\documentclass[aps,
twocolumn,reprint,
showpacs,preprintnumbers,
amsmath,amssymb,floatfix,a4paper,longbibliography]{revtex4-2} 

\usepackage{amsmath}
\usepackage{amsfonts}
\usepackage{amssymb}
\usepackage{bm}
\usepackage{bbm}
\usepackage{color}
\usepackage{graphicx}

\usepackage{hyperref}
\usepackage{dcolumn}
\usepackage{braket}
\usepackage{autobreak}
\usepackage{ulem}
\usepackage{xspace}
\usepackage{tikz}
\usetikzlibrary{quantikz2}
\usetikzlibrary{external}
\newcommand{\dquote}[1]{``#1''}

\newcommand{\etal}{\textit{et al.}}
\newcommand{\hc}{\mathrm{h.c.}}

\newcommand{\bs}{beam splitter\xspace}

\newcommand{\Hph}{H_{C}}
\newcommand{\Ppi}{P}
\newcommand{\nMode}{M}
\newcommand{\BTM}{B^\mathrm{TM}}
\newcommand{\PLJ}{P^{\mathrm{LJ}}}
\newcommand{\HLJ}{H^{\mathrm{LJ}}}
\newcommand{\btheta}{\bm{\theta}}
\newcommand{\UDD}{U_{\mathrm{DD}}}
\newcommand{\UDDrep}{\tilde{U}_{\mathrm{DD}}}

\newcommand{\CthreePO}{\mathrm{C3PO}}
\newcommand{\tphi}{T_{P}}
\newcommand{\Twait}{\tau}
\newcommand{\Thalf}{T_{\mathrm{50:50}}}
\newcommand{\Ca}{\mathrm{{}^{40}Ca^+}}

\newcommand{\acommu}{\tilde{\alpha}_{\mathrm{comm}}}
\newcommand{\erf}{\mathrm{erf}}
\newcommand{\Tup}{T_{\mathrm{u}}}
\newcommand{\Tdown}{T_{\mathrm{d}}}
\newcommand{\Tud}{T_{\mathrm{ud}}}
\newcommand{\NBP}{N_{\mathrm{BP}}}

\newcommand{\ii}{\mathrm{i}}

\newcommand{\mum}{\mathrm{\mu m}}

\newcommand{\mus}{\mathrm{\mu s}}

\newcommand{\kHz}{\mathrm{kHz}}
\newcommand{\MHz}{\mathrm{MHz}}
\newcounter{num}
\newcommand{\Rnum}[1]{\setcounter{num}{1} \Roman{num}}
\newcommand{\rnum}[1]{\setcounter{num}{1} \roman{num}}

\draft 

\begin{document}

\preprint{}

\title{Cancellation of phonon hopping in trapped ions by modulation of the trap potential} 

\author{Takanori Nishi}
\author{Kaoru Yamanouchi}
\email{kaoru@chem.s.u-tokyo.ac.jp}
\affiliation{Institute for Attosecond Laser Facility, The University of Tokyo, 7-3-1 Hongo, Bunkyo-ku, Tokyo 113-0033, Japan}
\author{Ryoichi Saito}
\author{Takashi Mukaiyama}
\affiliation{Department of Physics, Institute of Science Tokyo, 2-12-1 Ookayama, Meguro-ku, Tokyo 152-8550, Japan}
%


\begin{abstract}
The local modes of trapped ions can be used to construct an analog quantum simulator and a digital quantum computer. However, the control of the phonon hopping remains difficult because it proceeds among all the local modes through the Coulomb coupling.
We propose a method to cancel the phonon hopping among a given set of local modes by applying a sequence of phase shift gates implemented through the modulation of the trap potential.
We analyze the error scaling in the algorithm to treat three or more modes and show that the error can be suppressed by repeating the pulse sequence.
The duration of the phase shift gate in the present method can be as short as a few microseconds, which is an order of magnitude faster than the laser-based method. This short duration of the phase shift gate facilitates the suppression of the gate error.
We finally show how the present method can be applied to the implementation of the \bs.
The present method can also be applied to the simulation of bosonic systems as well as to the continuous variable encoding of quantum computing using trapped ions.

\end{abstract}


\maketitle 

\section{INTRODUCTION}\label{Sec:Intro}
The motional oscillator of trapped ions in a harmonic potential provides a controllable system of phonons, 
which can be used for continuous variable (CV) encoding of quantum information processing \cite{GKP2001, Braunstein2005} and quantum simulation of bosonic systems \cite{PorrasCirac2004BEC, Toyoda2013, KatzMonroe2023}.
In general, the Hamiltonian of the system composed of harmonic oscillators can be described either by the local mode or by the collective mode.
For the local mode, the Hamiltonian is composed of the on-site phonon interaction and the phonon hopping induced by the Coulomb repulsion,
while for the collective mode, the Hamiltonian matrix is obtained by diagonalizing the local-mode Hamiltonian matrix.

The local-mode scheme can be used to realize the quantum simulation of, e.g., a Bose--Hubbard model \cite{PorrasCirac2004BEC} and a Jaynes--Cummings--Hubbard model \cite{Toyoda2013}, 
and can also be used to implement a quantum walk \cite{Tamura2020QuantumWalk} and a boson sampling \cite{Shen2014}.
Moreover, it has been shown that a logical qubit in the CV encoding called the Gottesman--Kitaev--Preskill (GKP) state can be constructed using a local mode \cite{Fluhmann2019, deNeeve2022, MatsosPRL2024}.

However, in the local-mode scheme, the manipulation of entanglement among two or more local modes remains difficult because the phonon hopping, a source of entanglement, originating from the Coulomb interaction, cannot be turned off. 
On the other hand, in the collective mode scheme \cite{Chen2023}, the modes can be entangled using a beam-splitter gate, 
which can be switched on only by irradiating the ions with a laser \cite{Chen2023} or by applying the spatially and temporally varying electric potential \cite{Hou(NIST)2024TwoModeGate}.

There are several methods for the control of the phonon hopping.
In the explanation below, we refer to a system having two local modes for simplicity, but the discussion can be extended to a system having more than two local modes.
The first method is called the phonon blockade \cite{Ohira2021PhononBlockade}, where two local modes become off-resonant by the irradiation of one of the two local modes with laser light so that the phonon hopping can be suppressed.
The method protects the phonon state of one mode, but disturbs 
that of the other mode being excited by the laser light.

In the second method \cite{Lau&James2012}, the two ions are stored in distant traps where the phonon hopping is negligible, and they are transported into proximity by the deformation of the trap potential, kept for a certain time so that a desired amount of phonon hopping is induced, and then separated apart again.
This scheme requires a large-size ion storage area in which the distance between the two ions can be sufficiently long.

The third method is based on the idea of the dynamical decoupling (DD) \cite{Shen2014, Ohira2022}.
In this method, two modes are resonant and the phonon hopping proceeds from $t=0$ to $T/2$, then an instantaneous $\pi$-phase shift $\Ppi$ is applied to one mode, which flips the sign of the phonon hopping Hamiltonian $\Hph$, i.e., $\Hph\Ppi = -\Ppi\Hph$.
Therefore, the time evolution from $t=T/2$ to $T$ cancels with the phonon hopping in the first half period.
Because the phonon hopping proceeds even during the application of $\Ppi$, it is crucial to make the duration of $\Ppi$ negligibly short compared to the timescale of the phonon hopping.

In the original idea of the DD proposed by Shen \etal \cite{Shen2014}, the off-resonant sideband pulse was adopted to implement $\Ppi$.
However, when the practical experimental conditions are considered \cite{Ohira2022}, it is difficult to make the duration of $\Ppi$ sufficiently short because the rate of the off-resonant interaction is comparable to the hopping rate .
Ohira \etal \cite{Ohira2022} showed that a shorter $\pi$-phase shift $\Ppi$ can be realized by adopting the resonant sideband pulse, but their scheme can be used only when the phonon number is 0 or 1.

In the present study, we adopt the third method and propose to implement the $\pi$-phase shift by modulating the trap potential using an electric pulse applied to the trap electrodes and show that the fidelity of the DD can be improved.
The duration of $\Ppi$ can be a few microseconds, which is shorter by an order of magnitude than that which can be achieved by the resonant sideband pulse approach.
It should be noted that the present method can be applied to a system having any number of phonons.
We also show that the original DD approach ~\cite{Shen2014} needs to be modified when the number of modes becomes larger than two even when the duration of $\Ppi$ is negligibly small.
Finally, we show that the present method can be used to implement the \bs among the local modes.

\section{Method}\label{Sec:Theory}
\subsection{A sequence of phase shifts for the dynamical decoupling}
We consider a chain of ions and assume that the frequencies of the oscillation along the radial direction $\tilde{x}$, perpendicular to the trap axis, are a few MHz, while the hopping rate is a few kHz.
This condition allows us to adopt the local-mode picture along the $\tilde{x}$ axis. 
Therefore, if the number of trapped ions is $\nMode$, we consider $\nMode$ local modes and denote the set of those modes as $\mathcal{U}=\{q_j;0\leq j \leq \nMode-1\}$.
The scaled coordinate of the $j$-th mode $x_j$ is represented using the annihilation and creation operators as $x_j=(a_j^\dagger + a_j)/\sqrt{2}$.
As shown in Appendix~\ref{app:Hamiltonian}, when the distance between adjacent ions is much larger than the displacement of the respective ions from their equilibrium positions, the Hamiltonian is composed of the on-site interaction $H_0$ and the Coulomb interaction $\Hph$ as
\begin{align}
    H &= H_0 + \Hph \label{eq:Hamiltonian}\\
    H_0 &= \sum_{j=0}^{\nMode-1} \hbar\omega_0\left(a_j^\dagger a_j + \frac{1}{2}\right) \label{eq:H_0}\\
    \Hph &= \sum_{j>k}\frac{\hbar\kappa_{j,k}}{2}\left(a_j^\dagger + a_k^\dagger\right)\left(a_j + a_k\right) \notag\\
    &\simeq \sum_{j>k}\frac{\hbar\kappa_{j,k}}{2}\left(a_j^\dagger a_k + a_j a_k^\dagger\right)=\sum_{j>k} H_{j,k},\label{eq:Hph_RWA}
\end{align}
where $\omega_0$ in Eq. (2) is the secular frequency obtained under the influence of the Coulomb interactions.
The rotating-wave approximation (RWA) has been applied to $\Hph$ so that the fast oscillating terms such as $a_ja_k + a_j^\dagger a_k^\dagger$ can be omitted.
The Hamiltonian in the interaction picture with respect to $H_0$ is obtained as $H_I= \Hph$.
The time evolution under $H_I$ for the time duration $\tau$ is given by
\begin{align}
    \label{eq:U_Mmodes}
    &\exp(-\ii \tau H_I/\hbar) \notag\\
    &= B_M(\btheta_{\tau}) \equiv \exp\left(-\ii \tau \sum_{j>k}\frac{\kappa_{j,k}}{2} \left(a_j^\dagger a_k + a_j a_k^\dagger\right)\right),
\end{align}
where $B_{\nMode}$ is an $\nMode$-mode entangling gate and $(\btheta_{\tau})_{j,k}\equiv\tau\kappa_{j,k} / 2$.

Because $B_M$ induces the phonon hopping among all the modes, when we need to induce the phonon hopping among a set of modes $\mathcal{S}$ as
\begin{align}
    \label{eq:B_S}
    B_{\mathcal{S}}=\exp\left(-\ii \tau \sum_{\substack{j>k \\ j,k \in\mathcal{S}}}\frac{\kappa_{j,k}}{2} \left(a_j^\dagger a_k + a_j a_k^\dagger\right)\right),
\end{align}
we should cancel the phonon hopping among the other modes $q_r\in\mathcal{S}^c$ as well as that between $q_r$ and $q_s$ for $q_s\in\mathcal{S}$ and $q_r\in\mathcal{S}^c$. 
For example, in order to implement the 50:50 \bs between $q_j$ and $q_{j+1}$, we need to wait for $\Twait=\Thalf=(\pi/4) / (\kappa_{j+1,j}/2)$ while the phonon hopping involving $q_r\in\mathcal{S}^c$ should be canceled, i.e., the phonon hopping between $q_{r_1}$ and $q_{r_2}$ for $q_{r_1},q_{r_2}\in\mathcal{S}^c$, that between $q_j$ and $q_r$, and that between $q_{j+1}$ and $q_r$ should be canceled.
Because the waiting time $\Twait$ depends on the operation time of the gate applied to the modes in $\mathcal{S}$,
we consider the DD to cancel the phonon hopping occurring in the course of the time propagation from $t=0$ to arbitrary time $T$. 
First, we explain the basic idea of the DD for a two-mode case
and then extend the method to a general case with the larger number of modes.

\subsubsection{Two-mode DD}
When only two modes are involved, the time propagation for the time duration $\tau$ corresponds to a two-mode \bs $\BTM_{1,0}(\theta_{\tau})$ given by
\begin{align}
    \label{eq:U_beamsplitter}
    &B_2(\theta_{\tau})\notag\\
    &= \BTM_{1,0}(\theta_{\tau}) = \exp\left(-\ii \tau \frac{\kappa_{1,0}}{2} \left(a_1^\dagger a_0 + a_1 a_0^\dagger\right)\right),
\end{align}
where $\theta_{\tau}\equiv \tau\kappa_{1,0} / 2$.
As shown in Appendix~\ref{app:relations}, when a $\pi$-phase shift is defined by
\begin{align}
    \label{eq:phase_shift}
    \Ppi_j = \exp\left(-\ii \pi a_j^\dagger a_j\right),~ j=0~\mathrm{or}~1,
\end{align}
the sign of $H_I$ flips as $H_I\Ppi_j = -\Ppi_j H_I$, from which $\BTM_{1,0}(\theta)\Ppi_j=\Ppi_j \BTM_{1,0}(-\theta)$ can be derived.
Therefore, the phonon hopping occurring in the course of the time propagation from $t=0$ to $T$ represented by $\BTM_{1,0}(\theta_T)$ can be canceled by applying either $\Ppi_0$ or $\Ppi_1$ at $t=T/2$ and $T$ as
\begin{align}
    \label{eq:DD_2mode}
    \UDD=\Ppi_j \BTM_{1,0}(\theta_{T/2})\Ppi_j \BTM_{1,0}(\theta_{T/2}) = I, \, j=0~\mathrm{or}~1.
\end{align}
where $I$ denotes the identity operator. It should be noted that we assume here the duration of $\Ppi_j$ is negligibly short.
We can interpret Eq. \eqref{eq:DD_2mode} as the following sequence: (i) The phonon hopping proceeds for the first half period [$\BTM_{1,0}(\theta_{T/2})$], (ii) $\Ppi_j$ is applied to flip the sign of $H_I$, (iii) the phonon hopping proceeds with $-H_I$ in the latter half, which cancels the phonon hopping in the first half as
\begin{align}
    \label{eq:DD_2mode_explain}
    &\BTM_{1,0}(\theta_{T/2})\Ppi_j \BTM_{1,0}(\theta_{T/2}) \notag\\
    &\hspace{0mm} = \Ppi_j \BTM_{1,0}(-\theta_{T/2})\BTM_{1,0}(\theta_{T/2}) = \Ppi_j
\end{align}
and (iv) the $\pi$-phase shift is compensated by the second $\Ppi_j$ at the end by using $P_j^2=I$.

\subsubsection{DD for three or more modes} \label{subsubsec:Mmode_DD}
The DD for the general case with more than three modes, called the concatenation, was proposed in Ref.~\cite{Shen2014}.
Here, we give an algorithm to construct a pulse sequence for the concatenation.
In the first step, we consider a chain of $\nMode$ modes and group them into two sets.
Although the DD works for an arbitrary grouping, 
we choose, for simplicity, the two sets as the first half $S_0=\{q_j;~0\leq j \leq \lfloor\nMode/2 \rfloor - 1\}$ 
and the last half $S_1=\{q_j;~\lfloor \nMode/2 \rfloor \leq j \leq M-1\}$.
By applying $\Ppi$ to the modes in $S_1$ at $t=T/2$, the sign of $H_{j,k}$ for $\{j,k;~q_j\in S_0$,\, $q_k\in S_1\}$ flips and the phonon hopping between the two sets occurring in the later half period cancels with that which occurred in the first half period, i.e., $q_j\in S_0$ and $q_k\in S_1$ are decoupled.
Then, in the second step, we split $S_0$ into $S_{00}$ and $S_{10}$ in the same manner as above and apply $\Ppi$ to the modes in $S_{10}$ at $t=T/4$ and $3T/4$, which flips the sign of $H_{j,k}$ for $\{j,k;~q_j\in S_{00}$,\, $q_k\in S_{10}\}$.
Similarly, we split $S_1$ into $S_{01}$ and $S_{11}$, and apply $\Ppi$ to the modes in $S_{11}$ at $t=T/4$ and $3T/4$.

In general, in the $l$-th step for $l\geq2$, the subsets of the previous step are split into two, and 0 or 1 is attached to the left of the subscript:
(i) The subset of the previous step $S_{\alpha_{l-1}}$, where $\alpha_{l-1}$ is a binary number, is split into $S_{0\alpha_{l-1}}$ and $S_{1\alpha_{l-1}}$ so that the number of elements of the two subsets satisfies $|S_{0\alpha_{l-1}}|=|S_{1\alpha_{l-1}}|$ or $|S_{0\alpha_{l-1}}|=|S_{1\alpha_{l-1}}|-1$.
(ii) Then, $\Ppi$ is applied to the modes in $S_{1\alpha_{l-1}}$ at $t=(2p-1)T/2^l$ for $p=1,\,2,\,\dots,\,2^{l-1}$.
When $|S_{\alpha_{l-1}}|=1$, we just attach 0 to the left of the subscript.
This procedure is repeated until $l=l_{\max}=\lceil\log_2{\nMode}\rceil$ so that the number of elements in each subset becomes one.
At this step, each mode $q_j$ corresponds to a subset $S_{\alpha_j}$ and the binary number $\alpha_j$ is equal to the number $\Pi_{j}$ of $\Ppi$ applied to the $j$-th mode.

Finally, we need to compensate for the unwanted $\pi$-phase shift.
For the two-mode DD, $\Ppi_j$ in $\UDD$ can be moved to the left using the relation 
$\BTM_{1,0}(\theta)\Ppi_j=\Ppi_j \BTM_{1,0}(-\theta)$, as shown by Eq. \eqref{eq:DD_2mode_explain}.
For the $\nMode$-mode DD, we can move all the $\Ppi_j$ in $\UDD$ to the left in a similar manner so that we have $\Ppi_j^{\Pi_j}$, which is equal to $I$ when $\Pi_j$ is even and is equal to $\Ppi_j$ when $\Pi_j$ is odd.
Because $\Pi_j$ is odd if the $q_j$ is in $S_1$,
we apply $\Ppi$ to all the modes in $S_1$ at $t=T$.
Therefore, after applying this compensation procedure, the number of $\Ppi_j$ is given by $\Pi_{j}=\alpha_j+\delta_{1,r}$, where $r$ is the rightmost bit of $\alpha_j$ and $\delta_{1,r}$ is the Kronecker delta.

Because $\Ppi$ is not applied to $q_0$, which corresponds to the binary number $\alpha_0=00\dots0$, we can immediately implement $B_{\mathcal{S}}$ defined by Eq.~\eqref{eq:B_S} using the DD described above.
First, we define $\nMode=|\mathcal{S}^c|+1$ and choose a mode $q'$ from $\mathcal{S}$.
Then, by regarding $q'$ as the zeroth mode, we apply the $\nMode$-mode DD to the set of modes $\mathcal{S}'=\{q'\}\cup\mathcal{S}^c$.
Because the phonon hoppings among the modes in $\mathcal{S}^c$ and that between $q'$ and any mode in $\mathcal{S}^c$ are canceled and because the choice of $q'$ to which $\Ppi$ is not applied is arbitrary, the phonon hopping is not affected only among $\mathcal{S}$ by the DD and, consequently, $B_{\mathcal{S}}$ is securely realized.
An example for $\nMode=3$ will be shown in Sec. \ref{subsubsec:results_2modeBS}.

We note that the present algorithm also works even if the roles of the subscripts \dquote{0} and \dquote{1} are interchanged at each step, even though we assign the subscript \dquote{1} to the subsets to which $\Ppi$ is applied to obtain the simple relation between $\alpha_j$ and $\Pi_j$.
For example, at the $l$-th step, we can apply $\Ppi$ to the modes in $S_{0\alpha_{l-1}}$ instead of the modes in $S_{1\alpha_{l-1}}$.
As shown in Sec. \ref{subsubsec:results_3mode_idealDD}, this degree of freedom can be exploited to reduce the error in the DD.

The unitary operator $\UDD$ describing the DD consists of the repetition of the set of $B_M(\btheta_{\tau})$ and $\Ppi$ as Eqs.~\eqref{eq:DD_2mode} and \eqref{eq:3modeDD}.
The number $\NBP$ of this set in $\UDD$ is given by $\NBP=2^{l_{\max}}$.
In practice, considering the finite pulse duration $\tphi$ of $\Ppi$, $\NBP$ is upper bounded by $\NBP<T/\tphi$.
Therefore, when $T$ and $\tphi$ are given, the upper bound for $\nMode$ is obtained as 
$\lceil\log_2 \nMode \rceil<\log_2(T/\tphi)\Rightarrow$
$\nMode<2^{\lfloor \log_2(T/\tphi) \rfloor}$, which limits the applicability of the DD to a large $\nMode$.

The problem of the applicability of the DD to the large $\nMode$ can be addressed by truncating the Coulomb interaction at a certain mode separation $\eta\in\mathbb{Z}$, i.e.,
by omitting $H_{j,k}$ for $|j-k|>\eta$.
This truncation can be rationalized because the hopping rate scales as $\kappa_{j,k}\propto (d|j-k|)^{-3}$, where $d$ denotes the distance between two neighboring ions.

For example, when $\nMode=N\eta$ with $N\geq2$, we divide the set of $\nMode$ modes into $N$ subsets as $S_0$, $S_1,\,\dots$, $S_{N-1}$ so that each subset contains $\eta$ modes.
Because $|j-k|>\eta$ is satisfied for $\{j,k;~q_j\in S_J,\, q_k\in S_K\}$ if $|J-K|>1$, we do not need to consider the DD to decouple $S_J$ from $S_K$ for $|J-K|>1$.
Therefore, we first apply $\Ppi$ to the modes in the subsets with odd subscripts, $S_{2n+1}$, at $t=T/2$, which decouples the modes in the neighboring subsets, $q_j\in S_{2n}$ and $q_k\in S_{2(n+1)}$, from $q_l\in S_{2n+1}$.
Then, we apply the DD to each of $S_{2n}$, $S_{2n+1}$, and $S_{2(n+1)}$, so that $\NBP$ is given by $2^{\beta}$ with $\beta = \mathrm{min}(\lceil\log_2\eta\rceil+1,\, l_{\max})$.

\subsubsection{Repeated-DD} \label{subsubsec:repeated-DD}
As shown by Eq. \eqref{eq:DD_2mode}, the two-mode DD can completely cancel the phonon hopping, but the $M$-mode DD with $M\geq3$ cannot do so even if $\tphi=0$. Because we associate \dquote{1} in the subscript of $S_{\alpha_j}$ with the application of $\Ppi$, the unitary operator describing the three-mode DD is given by 
\begin{align}
    \label{eq:3modeDD}
    \UDD= P_2P_1B_3(\btheta_{\tau})P_2B_3(\btheta_{\tau})P_2P_1B_3(\btheta_{\tau})P_2B_3(\btheta_{\tau}),
\end{align}
where $\tau=T/\NBP$ and $\NBP=4$. 
By introducing the notation
\begin{align}
    \label{eq:3modeB+-}
    B_{\pm\pm\pm}&\equiv\exp\left[-\ii\frac{\tau}{2}
    \left(
    \pm\kappa_{1,0}a_1^\dagger a_0 
    \pm\kappa_{2,1}a_2^\dagger a_1
    \pm\kappa_{2,0}a_2^\dagger a_0 \right.\right. \notag\\
    &\left.\left.\hspace{60mm}+ \hc\right)\right],
\end{align}
and using the relations (see Appendix \ref{app:relations})
\begin{align}
    B_{\pm\pm\pm} P_1 = P_1 B_{\mp\mp\pm}, \label{eq:3mode_relation_P1_and_B+-}\\
    B_{\pm\pm\pm} P_2 = P_2 B_{\pm\mp\mp},\label{eq:3mode_relation_P2_and_B+-}
\end{align}
and $P_j^2=I$, Eq.~\eqref{eq:3modeDD} can be rewritten as
\begin{align}
    \label{eq:3modeDD_rewritten}
    \UDD&= P_2P_1B_{+++}P_2B_{+++}P_2P_1B_{+++}P_2B_{+++}\notag\\
    &= B_{-+-}B_{--+}B_{+--}B_{+++},
\end{align}
where $B_3=B_{+++}$.
Unlike the two-mode DD \eqref{eq:DD_2mode}, the four operators in the last line of Eq.~\eqref{eq:3modeDD_rewritten} do not cancel with each other
and, consequently, $\UDD=I$ cannot be satisfied.

We give a rough estimation of the error in the DD by applying the first-order product approximation \cite{Childs2021TrotterError} to $B_{\pm\pm\pm}$ as
\begin{align}
    \label{eq:Trotterization_of_B+-}
    B_{\pm\pm\pm}&=
    \BTM_{1,0}(\pm(\btheta_{\tau})_{1,0})
    \BTM_{2,1}(\pm(\btheta_{\tau})_{2,1})
    \BTM_{2,0}(\pm(\btheta_{\tau})_{2,0})\notag\\
    &\hspace{5mm} + O(\acommu T^2 \NBP^{-2}),
\end{align}
where 
$\acommu=\sum_{\xi,\xi'=\{(1,0),(2,1),(2,0)\}}(-\kappa_{\xi}\kappa_{\xi'}/4)\|[\alpha_{\xi}, \alpha_{\xi'}]\|$,
$\alpha_{\xi=(j,k)}=a_j^{\dagger}a_k + a_j a_k^{\dagger}$,
and $\|A\|\geq0$ stands for the spectral norm of $A$.
Due to the relations $[a_j, a_k]=0$ for all $\{j,k\}$  and $[a^{\dagger}_j, a_k]=0$ for $j\neq k$, $[\alpha_{(j,k)},\alpha_{(j',k')}]$ becomes nonzero only when $(j=j', k\neq k')$ or $(j\neq j', k=k')$ is satisfied.
Because both $\BTM_{j,k}(+(\btheta_{\tau})_{j,k})$ and $\BTM_{j,k}(-(\btheta_{\tau})_{j,k})$ appear twice and cancel with each other in the last line of Eq.~\eqref{eq:3modeDD_rewritten},
we can obtain $\UDD=I+ O(\acommu T^2 \NBP^{-2})$.
Furthermore, because the coupling strength $\kappa_{j+1,j}$ becomes the largest for the nearest-neighbor modes and is given by $\kappa_{j+1,j}=\kappa_{1,0}$ for an equidistant chain of ions, $\acommu$ can be approximated by $(-\kappa_{1,0}^2/4)\sum_{\xi,\xi'}\|[\alpha_{\xi}, \alpha_{\xi'}]\|$, where $\xi=(j+1, j)$ and $\xi'=(j'+1, j')$, and consequently, the error can be rewritten as $O(\kappa_{1,0}^2 T^2 \NBP^{-2})$.
It should be noted here that the time scale $T$ of the phonon hopping between the nearest-neighbor modes is proportional to $\kappa_{1,0}^{-1}$, e.g., the 50:50 \bs between the nearest-neighbor modes can be realized with $T=(\pi/4) / (\kappa_{1,0}/2)$.
Therefore, for inducing a fixed amount of phonon hopping, $\kappa_{1,0}^2$ cancels $T^2$ in the error term $O(\kappa_{1,0}^2 T^2 \NBP^{-2})$.
Moreover, because $\NBP$ is determined by the algorithm explained in Sec. \ref{subsubsec:Mmode_DD}, there is no free parameter which we can use to reduce the error in the DD during a fixed amount of phonon hopping.

In order to reduce the error, we replace $\tau$ in Eq.~\eqref{eq:3modeDD} by $\tau/n_r$ with a positive integer $n_r$, and repeat the pulse sequence $n_r$ times as
\begin{align}
    \label{eq:3modeDD_repeated}
    \UDDrep^{n_r} &=  \left\{P_2P_1
    B_3\left(\btheta_{\tau/n_r}\right)
    P_2
    B_3\left(\btheta_{\tau/n_r}\right) \right.\notag\\
    &\left.\hspace{6mm}
    P_2P_1
    B_3\left(\btheta_{\tau/n_r}\right)
    P_2
    B_3\left(\btheta_{\tau/n_r}\right) \right\}^{n_r},
\end{align}
which we call a repeated DD.
When $B_3$ is approximated by the product of $\BTM_{j,k}((\btheta_{\tau})_{j,k}/n_r)$ as in Eq.~\eqref{eq:Trotterization_of_B+-}, the error, which is scaled as in 
$\UDDrep^{n_r} =I+O(\acommu T^2 \NBP^{-2}n_r^{-1})$,  can be approximated to be $O(\kappa_{1,0}^2 T^2 \NBP^{-2}n_r^{-1})$.
Therefore, by increasing $n_r$, we can suppress the error and obtain $\UDDrep^{n_r}\simeq I$.

\subsection{Non-perturbative modulation of a harmonic potential}\label{subsec:c3po}
An off-resonant sideband pulse can be used to implement the $\pi$-phase shift, which can be applied to any number of phonons \cite{Shen2014}.
However, its interaction rate has been estimated to be less than $2.5\,\kHz$ in Ref.~\cite{Ohira2022}, which means that it would take $200\,\mus$ to achieve a $\pi$-phase shift.
The duration can be reduced to a few tens of microseconds using a resonant sideband pulse \cite{Ohira2022}, but this scheme is only applicable when the phonon number is 0 or 1.

Lau and James \cite{Lau&James2012} have shown that a phase shift which is faster and applicable to any number of phonons can be constructed by modulating the trap potential instead of irradiating the ions with laser light.
Here we describe this trap-potential modulation approach for a single harmonic oscillator whose Hamiltonian is given by $h_0=\hbar\omega_0 (a^\dagger a + 1/2)$.
As shown in Appendix~\ref{app:Lau&James}, when we apply a quadratic potential,
\begin{align}
    \label{eq:Vt}
    \HLJ(t) = \frac{\hbar\Omega^2(t)}{4\omega_0} \left(a^\dagger + a\right)^2,
\end{align}
the total Hamiltonian $h_0 + \HLJ(t)$ can be interpreted in the coordinate space as the harmonic oscillator oscillating with the time-dependent frequency $\omega(t)$, where $\Omega^2(t) = \omega^2(t) - \omega_0^2$.
When $\omega(t)$ is related with a real-valued and continuous function $b(t)$ as
\begin{align}
    \label{eq:def_b}
    \omega(t) = \sqrt{\left(\frac{\omega_0^2}{b^3} - \ddot{b}\right) / b},
\end{align}
and the boundary conditions $b(t\leq0)=b(t\geq\tphi)=1$ and $\omega(0)=\omega(\tphi)=\omega_0$ are satisfied,
the application of $\HLJ(t)$ from $t=0$ to $\tphi$ introduce a phase-shift operator, $\exp (-\ii\phi (a^\dagger a + 1/2))$, with the amount of the phase shift given by
\begin{align}
    \label{eq:phase_shift_LJ}
    \phi = \omega_0 \left(\int_0^{\tphi} \frac{dt}{b^2} - \tphi \right).
\end{align}
In addition, the $b$ function should be chosen so that the argument in the square root of the right-hand side of Eq.~\eqref{eq:def_b} is always non-negative and that $\ddot{b}$ is continuous.
We denote the $\pi$-phase shift induced by $\HLJ$ as $\PLJ$.
By replacing $\Ppi$ in $\UDD$ by $\PLJ$, we can cancel the Coulomb coupling by the modulation of harmonic potential. Hereafter, we call this approach the cancellation of Coulomb coupling by modulating harmonic potential (C3PO).
We note that $\PLJ=\exp (-\ii\pi (a^\dagger a + 1/2))$ differs from the conventional definition of the $\pi$-phase shift \eqref{eq:phase_shift} by the factor $\exp (-\ii\pi / 2)$, which only affects the global phase.
We adopt the $b$ function defined by using the error function as
\begin{widetext}
\begin{equation}
    \label{eq:b_errorfuncion}
    b=\left\{
    \begin{array}{cc}
         1-\frac{k}{2}\left(1+\erf\left[\left(\frac{t}{\Tup} -\frac{1}{2}\right)\sigma\right]\right)& (0\leq t\leq\Tup) \\
         1-k& (\Tup <t<\tphi-\Tdown) \\
         1-\frac{k}{2}\left(1-\erf\left[\left(\frac{t-(T-\Tdown)}{\Tdown} -\frac{1}{2}\right)\sigma\right]\right)& (\tphi-\Tdown\leq t\leq \tphi)
    \end{array}
    \right.,
\end{equation}
\end{widetext}
where $\Tup$ and $\Tdown$ are the ramp-up and the ramp-down time of the potential modulation, respectively. 
The width of the error function $\sigma$ should be large enough to ensure that $b$ is smooth at $t=\Tup$ and $T-\Tdown$.
In the present study, we fix $\sigma=6$ and $\Tup=\Tdown=\Tud$.
For a given pulse duration $\tphi$, the strength $k$ is determined so that Eq.~\eqref{eq:phase_shift_LJ} is satisfied for $\phi=\pi$ (see Appendix \ref{app:Lau&James}).

Because the secular frequency of an ion trapped in a radio-frequency (RF) trap is dependent on the amplitude of the RF and the DC voltages \cite{Wineland1998}, the time-dependent frequency $\omega(t)$ can be implemented by the modulation of the RF and/or DC voltage(s). 
The relation between $\omega(t)$ and the RF/DC amplitudes for a linear trap is given in Appendix \ref{app:electric_pulse}.
However, it is not an easy task to control the trap potential for one ion while keeping the potentials for the other ions the same. Using a surface electrode trap, it was shown that the secular frequency of a local mode can be adjusted without affecting the other modes by applying a specific set of DC voltages \cite{Mielenz2016}.
Therefore, it is expected that the C3PO, which requires the application of $\PLJ$ by individually modulating the trap frequency, can be implemented using a surface electrode trap.

\section{Results} \label{sec:Results}
We demonstrate the C3PO by numerically solving the time-dependent Schr\"odinger equation (TDSE) from $t=0$ to $T$,
where $T$ is set to the period of the 50:50 \bs between the neighboring modes.
In order to solve the TDSE efficiently and accurately, the semiglobal method \cite{Schaefer2017SemiGlobal} has been adopted.
There are two sources of error in the C3PO: (i) the algorithmic error due to the imperfect cancellation of the phonon hopping as explained in Sec. \ref{subsubsec:Mmode_DD} and (ii) the gate error due to the finite pulse duration of $\PLJ$.
Because the algorithmic error is absent for $\nMode=2$,
we first examine the two-mode C3PO to determine the optimal parameters which minimize the gate error.
Then, both the algorithmic and the gate error are investigated for $\nMode=3$.
Hereafter, when the instantaneous $\pi$-phase shift is applied instead of $\PLJ$, we call the DD the ideal DD.
We evaluate the error by $\braket{E}\equiv|\bra{\psi_0}I-U\ket{\psi_0}|=1-|\bra{\psi_0}U\ket{\psi_0}|$ for an initial state $\ket{\psi_0}$, where $U$ represents the pulse sequence of the ideal DD or the C3PO.

\subsection{2-mode C3PO} \label{subsec:results_2mode_C3PO}
We consider a chain of $\Ca$ ions.
The oscillation frequency along $\tilde{x}$ is set to $\omega_0/(2\pi) = 2.2\,\MHz$.
The state of $\nMode$ phonons is represented as $\ket{\psi}=\sum_j c_j\ket{j}$, where $\ket{j}=\ket{n_{\nMode-1}, \dots,\,n_1,\,n_0}$ and $n_m$ is the number of phonons in the $m$-th mode.

For $\nMode=2$, the initial state is chosen as $\ket{\psi_0}=\ket{2,\,1}$ so that the applicability of the C3PO to the cases with more than one phonon is demonstrated.
The pulse sequence of the C3PO can be represented by $U_{\CthreePO}=\PLJ_1 \BTM_{1,0}(\theta_{T/2})\PLJ_1 \BTM_{1,0}(\theta_{T/2})$, 
where $T=\Thalf=(\pi/4) / (\kappa_{1,0}/2)$.
We examine the effect of the Coulomb coupling by changing the pulse duration $\tphi$ and the distance between neighboring ions $d$.
First, we set $d=27.6\,\mum$ so that the hopping rate becomes the same as that in the previous study \cite{Ohira2022}, $\kappa_{1,0}/(2\pi)=1.9\,\kHz$.

\begin{figure}[htbp]
\begin{center}
\includegraphics[width=8.5cm]{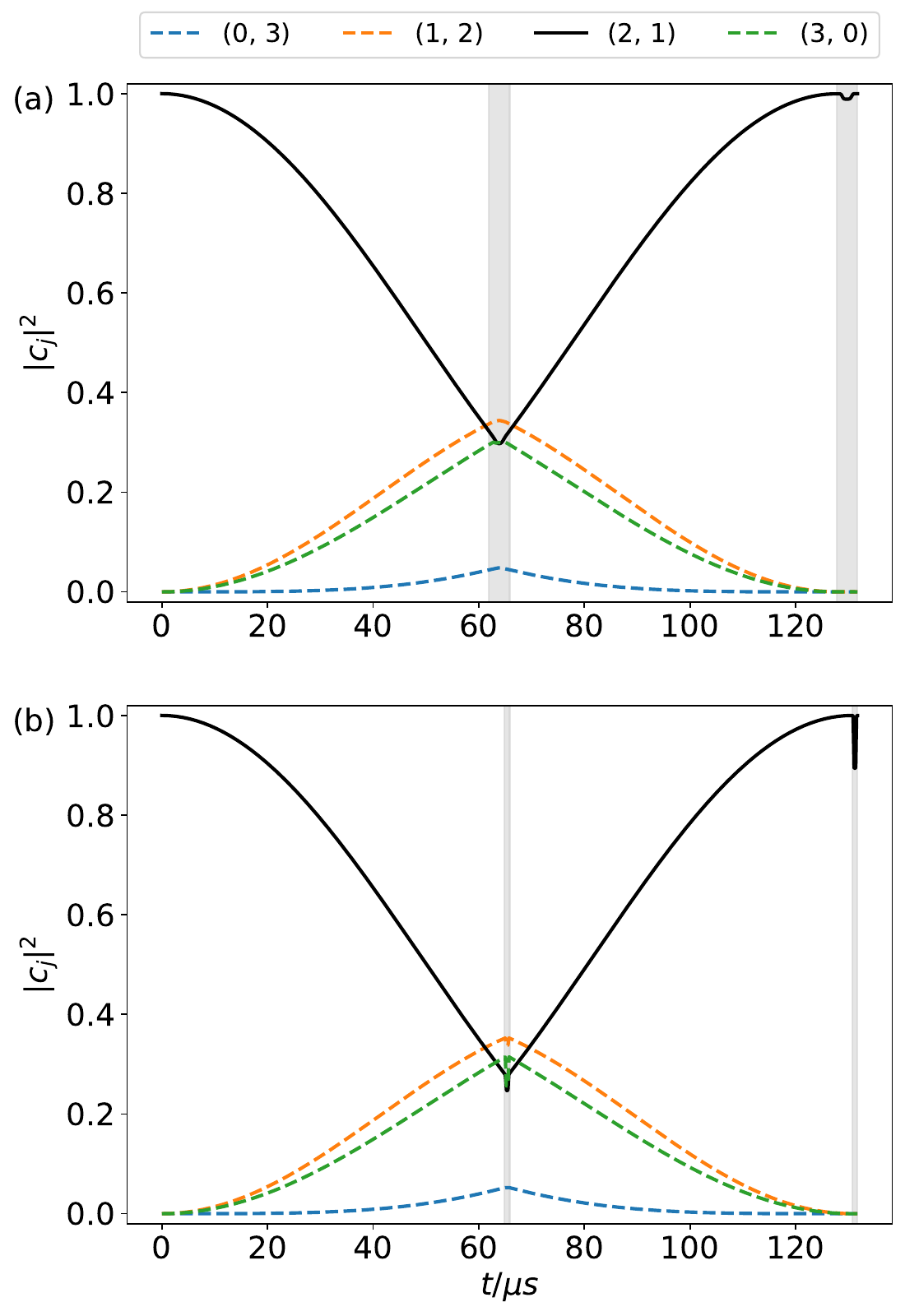}
\caption{The populations of a two-mode phonon state during the C3PO with $d=27.6\,\mum$.
(a) $\tphi=4.0\,\mus$ and (b) $\tphi=1.0\,\mus$.
The initially occupied phonon state is shown by the solid curve, while the other states are shown by the dashed curves.
The application of $\PLJ$ is indicated by the vertical shaded areas.}
\label{fig:2mode_C3PO_d27.6T4andT1}
\end{center}
\end{figure}

As shown in Appendix \ref{app:Lau&James}, when we set $\tphi=4\,\mus$, the secular frequency increases from $2\pi\times2.2\,\MHz$ by $2\pi\times250\,\kHz$ in the first $2\,\mus$.
Because it was shown in Ref.~\cite{Mielenz2016} that the secular frequency increases from $2\pi\times2.6\,\MHz$ by $2\pi \times 430\,\kHz$ by ramping up the DC voltage in $7.5\,\mus$,
we expect that $\PLJ$ can be realized with $\tphi=4\,\mus$.  
The result of the simulation of $U_{\CthreePO}\ket{\psi_0}$ with $\tphi=4\,\mus$ is shown in Fig.~\ref{fig:2mode_C3PO_d27.6T4andT1}(a).
The population of $\ket{2,\,1}$ (solid curve) decreases until $t=T/2$ and transfers to the other states (dashed curves), which conserves the total number of phonons.
The time evolution in the first half $0\leq t\leq T/2$ is canceled by the latter half $T/2\leq t \leq T$, and the resultant error is $\braket{E}=1.0\times10^{-5}$.

\begin{figure}[htbp]
\begin{center}
\includegraphics[width=8.5cm]{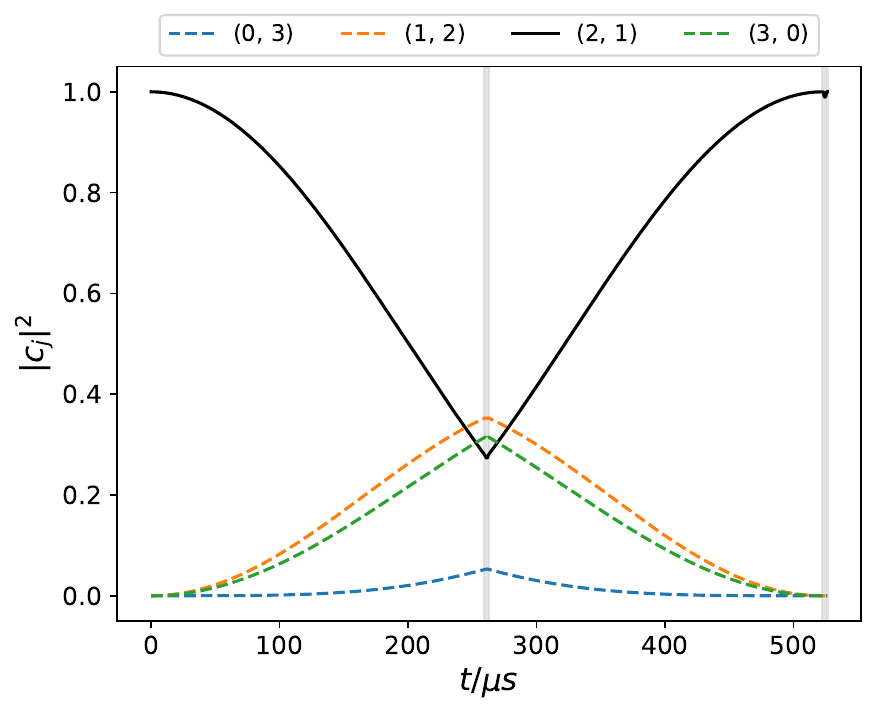}
\caption{The populations of a two-mode phonon state during the C3PO.
$\tphi=4.0\,\mus$ and $d=43.8\,\mum$.
The initially occupied phonon state is shown by a solid curve, while the other states are shown by dashed curves.
The application of $\PLJ$ is indicated by the vertical shaded areas.}
\label{fig:2mode_C3PO_d43.8T4}
\end{center}
\end{figure}

We can reduce the error by shortening the pulse duration or/and slowing down the phonon hopping.
For example, as shown in Fig.~\ref{fig:2mode_C3PO_d27.6T4andT1}(b), when $\tphi$ is set to $1\,\mus$, the error is lowered to be $\braket{E}=2.2\times10^{-6}$.
Instead of making $\tphi$ four times smaller, which requires an improvement in the experimental setup, we may make the phonon hopping slower by four times, which can be easily achieved by replacing $d$ with $4^{(1/3)}d$.
As shown in Fig.~\ref{fig:2mode_C3PO_d43.8T4}, for $d=43.8\,\mum$, which corresponds to $\kappa_{1,0}/(2\pi)=0.47\,\kHz$, the error is lowered further to $\braket{E}=2.2\times10^{-8}$.

Although we only show the populations of the states that preserve the number of phonons,
other states can also be populated during the application of $\PLJ$.
This can be most clearly seen in the case of $\tphi=1\,\mus$.
The expanded view of the shaded area of Fig.~\ref{fig:2mode_C3PO_d27.6T4andT1}(b) is given in Fig.~\ref{figApp:zoom} in Appendix \ref{app:Lau&James}.

\subsection{3-mode C3PO} \label{subsec:results_3mode_C3PO}
Here we adopt, for the three-mode case ($\nMode=3$), the same parameters that we adopted when obtaining the result shown in Fig.~\ref{fig:2mode_C3PO_d43.8T4}, i.e., $\tphi=4.0\,\mus$ and $d=43.8\,\mum$, which are experimentally achievable. With the set of parameters, we expect that the error can be sufficiently small.
As shown below, $\ket{\psi_0} = \ket{2,\,1,\,0}$ is chosen as the initial state in the numerical simulations of the cancellation of the phonon hopping, while $\ket{\psi_0} = \ket{1,\,1,\,1}$ is chosen as the initial state in the simulation of the \bs.

\subsubsection{Ideal DD} \label{subsubsec:results_3mode_idealDD}
\begin{figure}[htbp]
\begin{center}
\includegraphics[width=8.5cm]{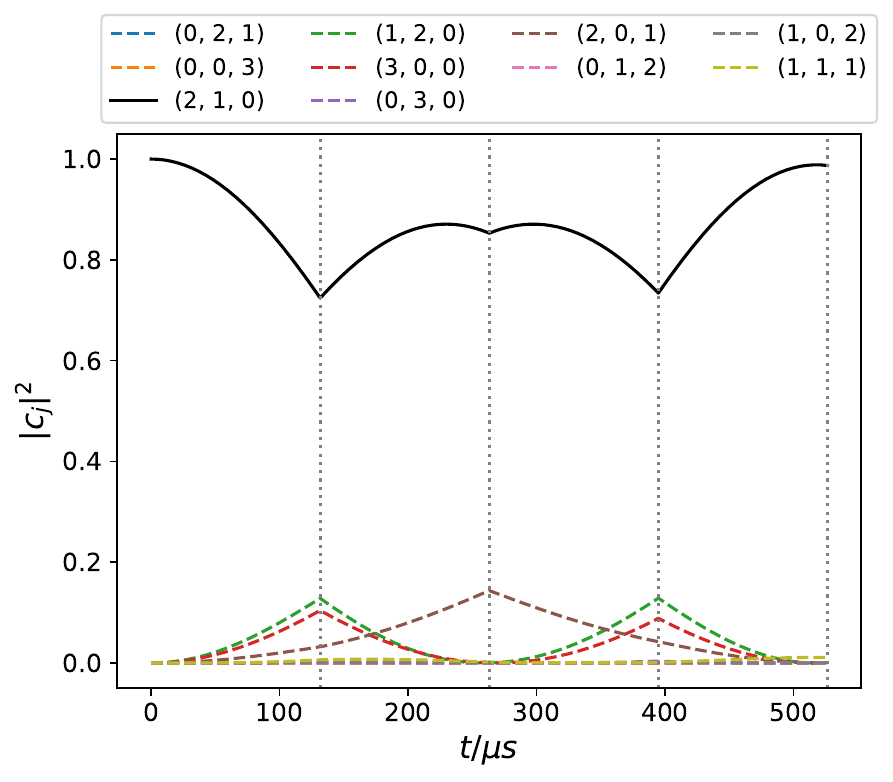}
\caption{The populations of a three-mode phonon state during the single cycle of the ideal DD. 
The pulse sequence is defined by Eq. \eqref{eq:3modeDD}.
The initially occupied phonon state is shown by a solid curve, while the other states are shown by dashed curves.
The application of the instantaneous $\pi$-phase shifts is indicated by the vertical dotted lines.}
\label{fig:3modeDD_1rep_2-21-2-21}
\end{center}
\end{figure}

A single cycle of the ideal DD for $T=\Thalf=(\pi/4) / (\kappa_{j,j-1}/2),\, j=1\,\mathrm{or}\,2,$ is illustrated in Fig.~\ref{fig:3modeDD_1rep_2-21-2-21}.
The unitary operator representing the time evolution is given by Eq.~\eqref{eq:3modeDD} and the error is obtained to be $\braket{E}=6.4\times10^{-3}$.
As described in Sec. \ref{subsubsec:Mmode_DD}, the sequence of the $\Ppi$ pulses is not uniquely determined.
Here we introduce another type of pulse sequence, which can make the error smaller:
(i) We split the three modes into two subsets $S_0=\{q_0\}$ and $S_1=\{q_1,\,q_2\}$ and apply $\Ppi$ to the modes in $S_1$ at $t=T/2$.
(ii) We split $S_1$ into $S_{01}=\{q_1\}$ and $S_{11}=\{q_2\}$ and apply $\Ppi$ to the modes in $S_{01}$ [unlike Eq. \eqref{eq:3modeDD}, where $\Ppi$ is applied to the modes in $S_{11}$] at $t=T/4$ and $3T/4$.
(iii) Finally, the unwanted $\pi$-phase shift is canceled by applying $\Ppi$ to the modes in $S_1$ at $t=T$.
The unitary operator describing this sequence is given by
\begin{align} 
    \UDD&= P_2P_1B_3(\btheta_{\tau})P_1B_3(\btheta_{\tau})P_2P_1B_3(\btheta_{\tau})P_1B_3(\btheta_{\tau})\label{eq:UDD_1_12_1_12}\\
    &=B_{-+-}B_{+--}B_{--+}B_{+++},
\end{align}
and the error is reduced to $\braket{E}=4.4\times10^{-5}$.
This significant improvement shows the importance of the optimization of the pulse sequence, which will be discussed further in a future work.

\begin{figure}[htbp]
\begin{center}
\includegraphics[width=8.5cm]{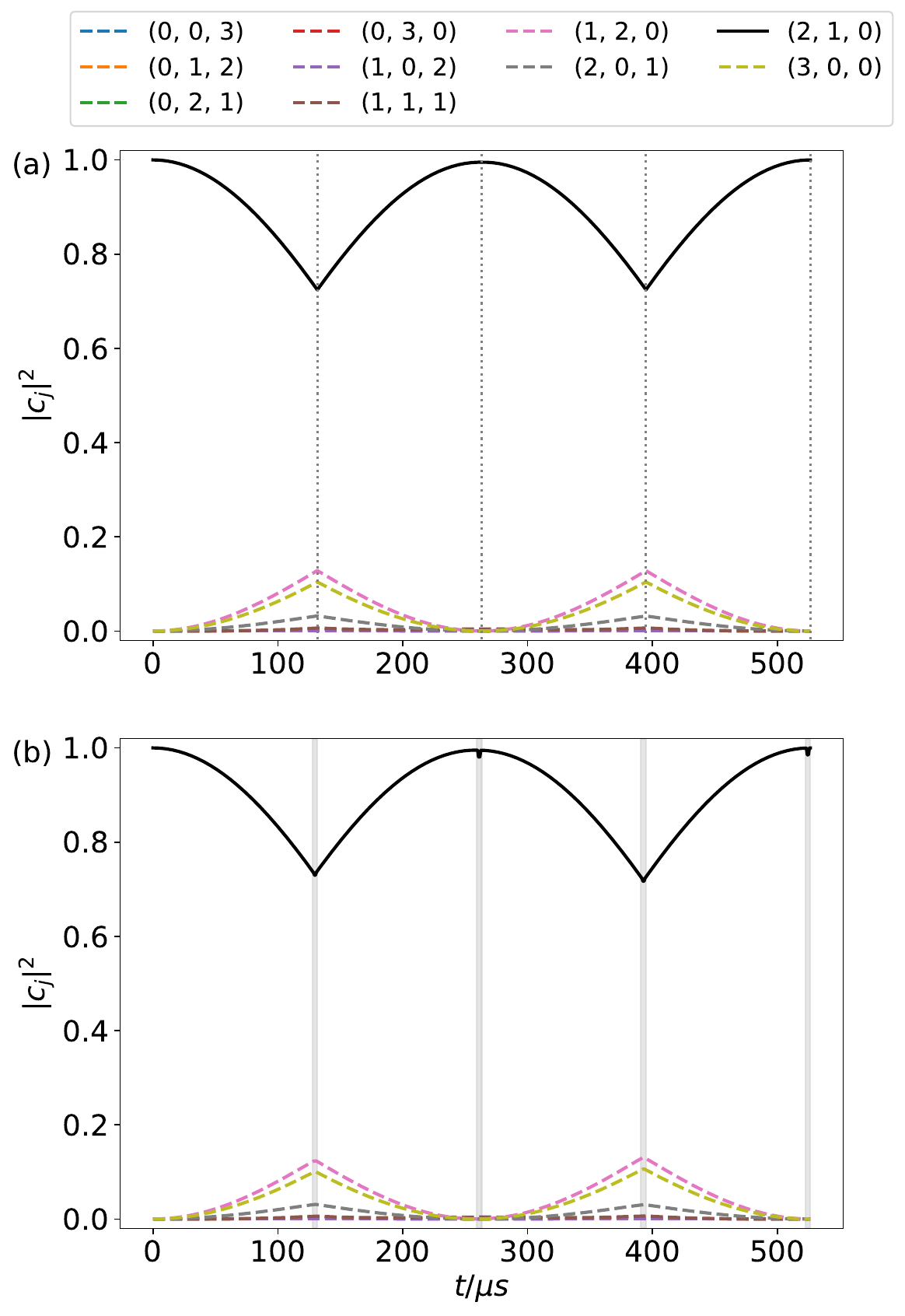}
\caption{The populations of a three-mode phonon state during the single cycle of (a) the ideal DD and (b) the C3PO. 
The pulse sequence is defined by Eq. \eqref{eq:UDD_1_12_1_12}.
The initially occupied phonon state is shown by a solid curve, while the other states are shown by dashed curves.
The application of the instantaneous $\pi$-phase shifts is indicated by the vertical dotted lines in (a),
while the application of $\PLJ$ is indicated by the vertical shaded areas in (b).}
\label{fig:3modeDD_1rep}
\end{center}
\end{figure}

Although the pulse sequence is designed so that the initial state is recovered at $t=T$, the initial state is almost recovered at $t=T/2$  in Fig.~\ref{fig:3modeDD_1rep}(a), and the dynamics from $t=0$ to $T/2$ is repeated from $t=T/2$ to $T$.
This is because $P_1$ applied at $t=T/4$ flips the sign of the nearest-neighbor coupling terms, $a_1^\dagger a_0 +\hc$ and $a_2^\dagger a_1 + \hc$, which have the dominant contribution to the dynamics,
while $P_2$ applied at $t=T/2$ flips the sign of the non-nearest-neighbor coupling term $a_2^\dagger a_0 + \hc$, whose contribution is eight times smaller in magnitude because $\kappa_{2,0}\propto (2d)^{-3}$.
This observation supports the validity of the truncation described in Sec. \ref{subsubsec:Mmode_DD}, where the pulse sequence is constructed to cancel the hopping between ions in the distance smaller than or equal to $\eta d$.

\begin{figure}[htbp]
\begin{center}
\includegraphics[width=8.5cm]{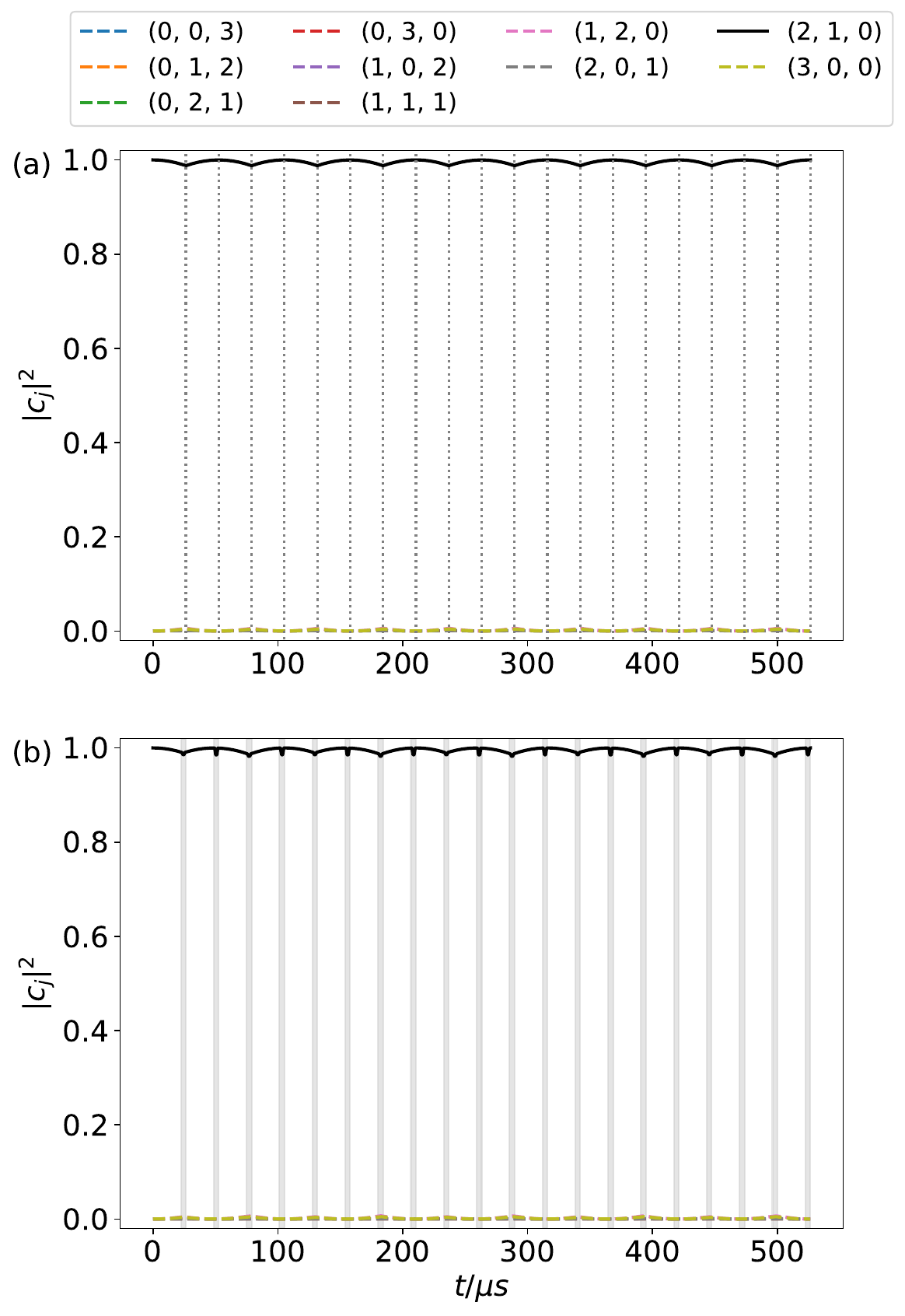}
\caption{The populations of a three-mode phonon state during the five cycles of (a) the ideal DD and (b) the C3PO. 
The pulse sequence is obtained by repeating $\UDD$ defined by Eq. \eqref{eq:UDD_1_12_1_12}.
The initially occupied phonon state is shown by a solid curve, while the other states are shown by dashed curves.
The application of the instantaneous $\pi$-phase shifts is indicated by the vertical dotted lines in (a),
while the application of $\PLJ$ is indicated by the vertical shaded areas in (b).}
\label{fig:3modeDD_5rep}
\end{center}
\end{figure}

The ideal repeated DD $\UDDrep^{n_r}$ derived from Eq. \eqref{eq:UDD_1_12_1_12} can be similarly defined as in Eq.~\eqref{eq:3modeDD_repeated} and an example with $n_r=5$ is illustrated in Fig.~\ref{fig:3modeDD_5rep}(a).
We can see the five cycles of phonon hopping and its cancellation. 
We note that the error scaling given in Sec. \ref{subsubsec:repeated-DD} is not tight and the actual scaling for fixed $\kappa_{1,0}$ and $\NBP$ is estimated as $O(T^4 n_r^{-2})$ from the numerical simulation of the three-mode DD.
Here, the error for the repeated DD with $n_r=5$ is calculated to be $\braket{E}=1.9\times 10^{-6}$, which is about $n_r^2 = 25$ times smaller than the error obtained with $n_r=1$.

\subsubsection{C3PO} \label{subsubsec:results_3mode_C3PO}

By replacing $\Ppi_j$ in Eq. \eqref{eq:UDD_1_12_1_12} by $\PLJ_j$, a numerical simulation of the three-mode C3PO was performed with $n_r=1$ and $n_r=5$. The results of the simulation with $n_r=1$ and $n_r=5$ are shown in Figs.~\ref{fig:3modeDD_1rep}(b) and \ref{fig:3modeDD_5rep}(b), respectively.
The errors obtained for $n_r=1$ and $n_r=5$ are $\braket{E}=4.4\times 10^{-5}$
and $\braket{E}=2.6\times 10^{-6}$, respectively.

Although the algorithmic error can be reduced further by increasing $n_r$ in the case of the ideal DD, the upper limit of $n_r$ given using $\NBP n_r < T/\tphi$ exists for the C3PO due to the finite pulse duration.
In the current case with $\NBP=4$ and $T/\tphi \simeq 132$, we obtain $n_r < 33$.
By using $\NBP=2^{\lceil \log_2\nMode\rceil}$, the upper bound of $\nMode$ for a fixed $n_r$ is given as 
$\nMode < 2^{\lfloor\log_2(T/(\tphi n_r))\rfloor}$.
For example, $\nMode<128$ for $n_r=1$ and $\nMode<16$ for $n_r=5$.
Therefore, in order to apply the C3PO to a system with a large $\nMode$, we need to adopt the truncation of $H_{j,k}$ for $|j-k|>\eta$.
By requiring $N_{\nMode}=2^{\lceil\log_2\eta\rceil + 1}<T/(\tphi n_r)$, we obtain 
$\eta<2^{\lfloor\log_2(T/(\tphi n_r))\rfloor - 1}$, which means that, for example,
$\eta$ needs to be $\eta<64$ for $n_r=1$ and $\eta<8$ for $n_r=5$.

\subsubsection{Application of C3PO to the two-mode \bs} \label{subsubsec:results_2modeBS}
Here we implement the two-mode \bs by applying the C3PO to cancel unwanted phonon hoppings by the method described in Sec. \ref{subsubsec:Mmode_DD}.
We choose $\ket{\psi_0}=\ket{1,1,1}$ as the initial state and consider the \bs between $q_0$ and $q_1$.
First, we denote the subset to which the entangling operation defined by Eq. \eqref{eq:B_S} is applied as $\mathcal{S}$ and the rest of the modes as $\mathcal{S}^c$.
Then, the pulse sequence of the DD is constructed so that $\Ppi$ is not applied to $q'\in\mathcal{S}$,  where  $\mathcal{S}=\{q_0,\,q_1\}$ and $\mathcal{S}^c=\{q_2\}$.
This is equivalent to considering the two-mode DD between $q'$ and $q_2$.
Therefore, we apply $\Ppi$ to $q_2$ at $t=T/2$ and $t=T$.

\begin{figure}[htbp]
\begin{center}
\includegraphics[width=8.5cm]{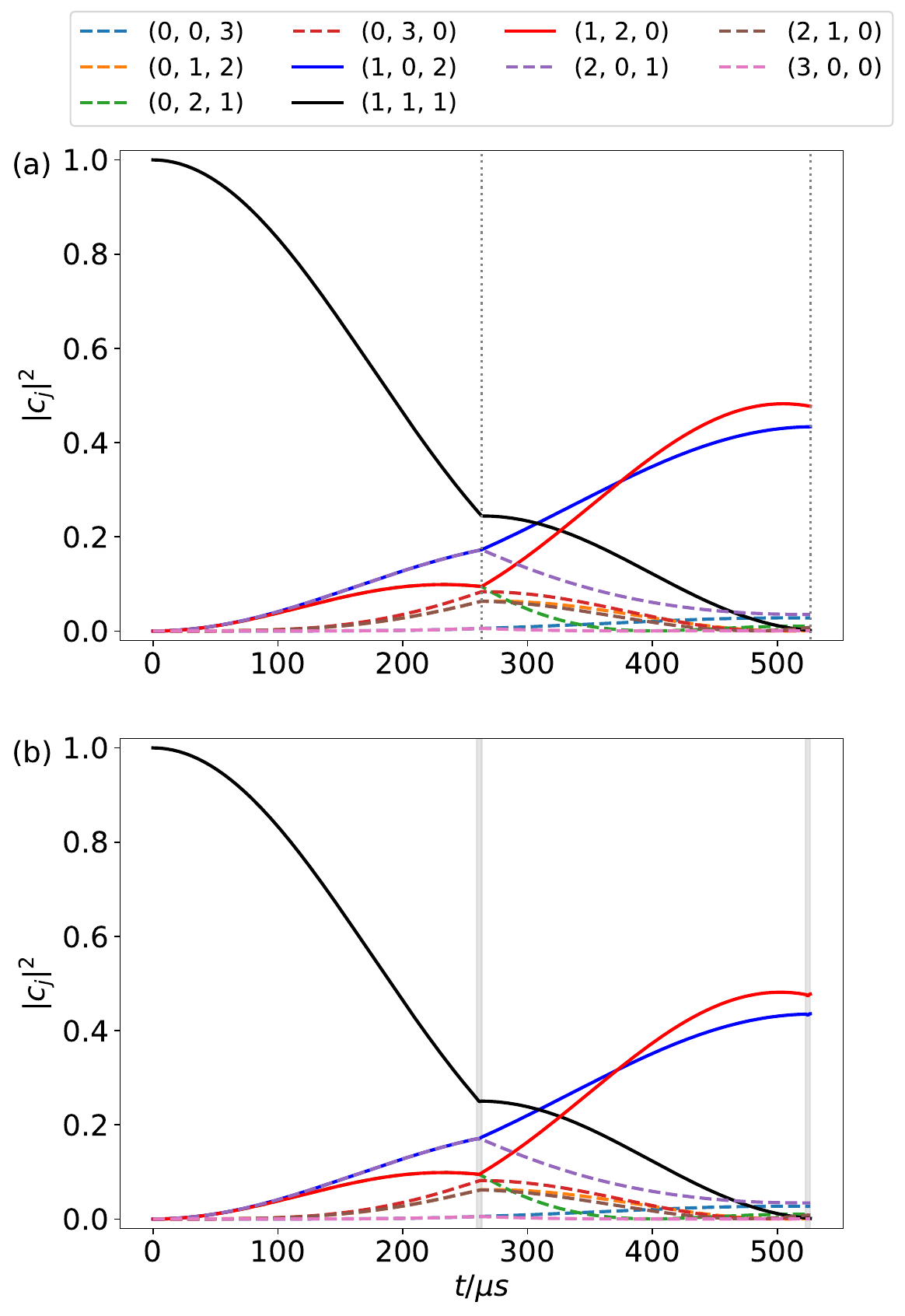}
\caption{The populations of a three-mode phonon state during the two-mode \bs implemented by (a) the ideal DD and (b) the C3PO. 
The initial state $\ket{1,1,1}$ (black) and the desired final states, $\ket{1,2,0}$ (red) and $\ket{1,0,2}$ (blue), are shown by solid curves, while the other states are shown by dashed curves.
The application of the instantaneous $\pi$-phase shifts is indicated by the vertical dotted lines in (a),
while the application of $\PLJ$ is indicated by the vertical shaded areas in (b).}
\label{fig:2modeBS_1rep}
\end{center}
\end{figure}

\begin{figure}[htbp]
\begin{center}
\includegraphics[width=8.5cm]{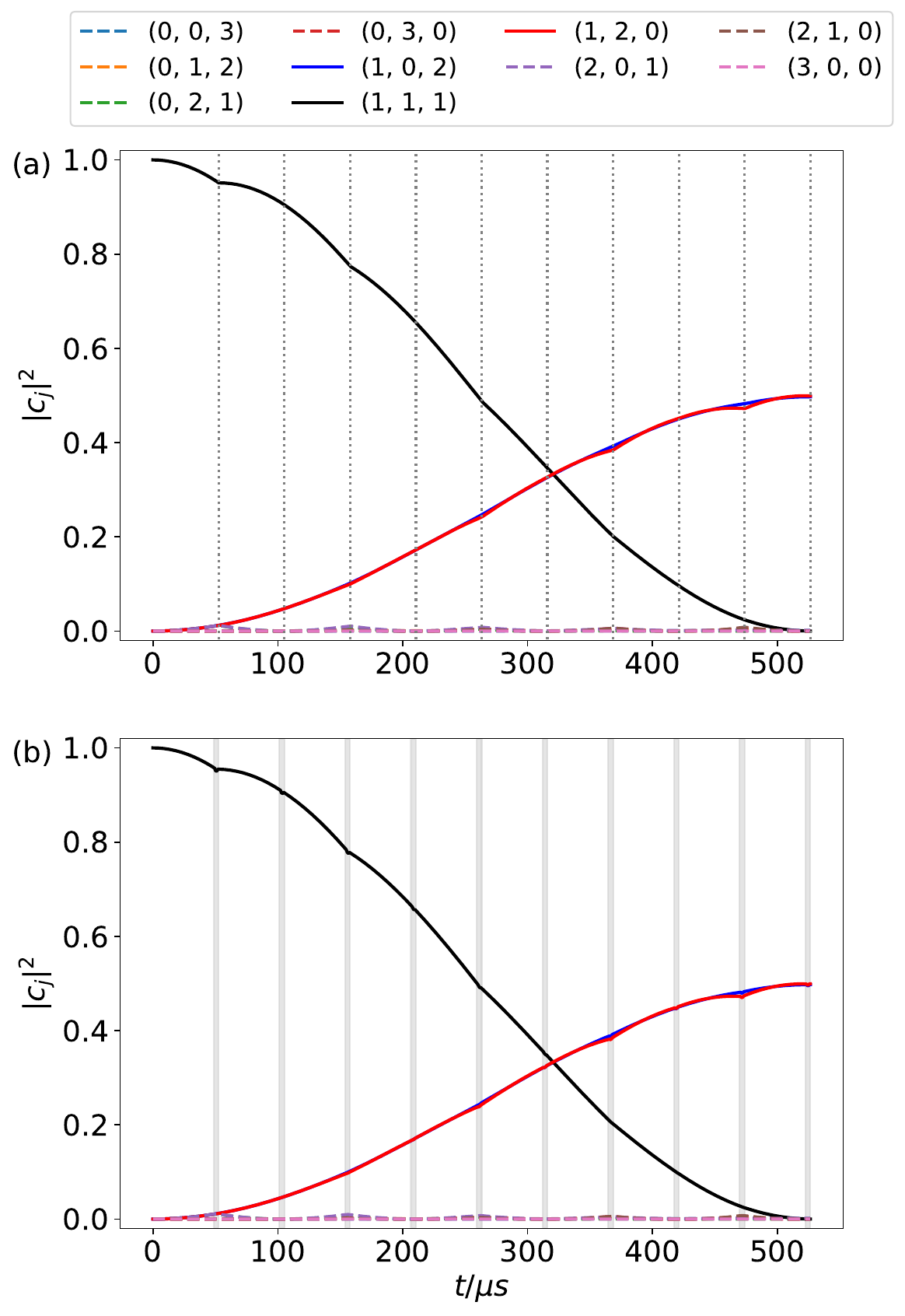}
\caption{The populations of a three-mode phonon state during the two-mode \bs implemented by the five cycles of (a) the ideal DD and (b) the C3PO. 
The initial state $\ket{1,1,1}$ (black) and the desired final states, $\ket{1,2,0}$ (red) and $\ket{1,0,2}$ (blue), are shown by solid curves, while the other states are shown by dashed curves.
The application of the instantaneous $\pi$-phase shifts is indicated by the vertical dotted lines in (a),
while the application of $\PLJ$ is indicated by the vertical shaded areas in (b).}
\label{fig:2modeBS_5rep}
\end{center}
\end{figure}

The results of the simulation with $n_r=1$ and $n_r=5$ are shown in Figs. \ref{fig:2modeBS_1rep} and \ref{fig:2modeBS_5rep}, respectively.
Because the input to the 50:50 \bs is $\ket{1,1}$, the output should be $(\ket{2,0}+\ket{0,2})/\sqrt{2}$ due to the Hong--Ou--Mandel effect \cite{HongOuMandel1987, Toyoda2015}.
Therefore, the final state is $\ket{\psi_f}=(\ket{1,2,0}+\ket{1,0,2})/\sqrt{2}$, where both the populations of $\ket{1,2,0}$ and $\ket{1,0,2}$ increase and become close to 0.5 as expected.
The error is defined as 
$\braket{E_B} \equiv |\bra{\psi_f} U_{\mathrm{ideal}}-U \ket{\psi_0}| = 1-|\bra{\psi_f} U \ket{\psi_0}|$, where $U_{\mathrm{ideal}}=I_2\otimes B_{1,0}(\Thalf)$ is the desired unitary operator implementing the 50:50 \bs between $q_1$ and $q_0$, $\ket{\psi_f}=U_{\mathrm{ideal}}\ket{\psi_0}$ is the desired final state, and $U$ represents the pulse sequence of the ideal DD or the C3PO.
When $n_r=1$, the error is obtained to be $\braket{E_B}=4.6\times10^{-2}$ for the ideal DD and $\braket{E_B}=4.5\times10^{-2}$ for the C3PO, 
and when $n_r=5$, the error can be suppressed to be $\braket{E_B}=1.8\times10^{-3}$ for the ideal DD and $\braket{E_B}=1.6\times10^{-3}$ for the C3PO.

\section{Conclusion}
We have proposed a method called C3PO to control phonon hopping among the local modes in a chain of trapped ions by applying a sequence of phase-shift operations realized by the modulation of the trap potential.
The present method can be applied to a system having any number of phonons and modes, which has been demonstrated by the numerical simulations of the cancellation of the phonon hopping among local modes.
By applying the C3PO to cancel the phonon hopping among all the modes except the chosen set of modes $\mathcal{S}$, we can implement the multi-mode entangling gate for $\mathcal{S}$.
We have shown that the two-mode \bs can be implemented between a set of two modes chosen from a chain of three modes by canceling the unwanted phonon hopping by the C3PO.
It should be noted that the C3PO can be applied to the boson sampling \cite{Shen2014} and the logical two-qubit gate for the GKP state \cite{Rojkov2024}.
For the experimental demonstration of the C3PO, we need to examine possible sources of error such as imperfect control of the phase-shift pulse, motional heating, and vibrational decoherence.
We note that the logical two-qubit gate for the GKP states encoded using two radial modes of a single ion was recently demonstrated \cite{Matsos2024}, 
but this approach can be applied to a system composed of three or less modes because it relies on the coupling between the electronic and vibrational degrees of freedom of a single ion.

\begin{acknowledgments}
This work was supported by JST-CREST Grant No. JPMJCR23I7.
\end{acknowledgments}

\appendix
\section{Hamiltonian}\label{app:Hamiltonian}
We consider $\nMode$ trapped ions oscillating along the $\tilde{x}$ axis.
The Hamiltonian of the $j$-th ion is given by 
$H_j = \tilde{p}_j^2/(2m) + m\tilde{\omega}_0^2 \tilde{x}_j^2/2$,
where $m$ is the ion's mass and $\tilde{\omega}_0$ is the oscillation angular frequency.
The Coulomb interaction between two cations located at positions $\bm{r}_1$ and $\bm{r}_2$ can be approximated as \cite{Rojkov2024}
\begin{align}
    \label{eqApp:coulomb}
    \frac{e^2}{4\pi\epsilon_0}\frac{1}{|\bm{r}_1-\bm{r}_2|}
    \simeq \frac{e^2}{4\pi\epsilon_0}\left(\frac{1}{d_{1,2}} + \frac{-\tilde{x}_1^2 - \tilde{x}_2^2 + 2\tilde{x}_1\tilde{x}_2}{2d_{1,2}^3}\right),
\end{align}
where $d_{1,2}$ is the distance between the two equilibrium positions of the two cations. In deriving Eq. (A1), we use the fact that the displacement of each ion from the equilibrium position is perpendicular to the trap axis.
Hereafter, we neglect the first term in the right-hand side of Eq. (A1) because it only affects the global phase.
Then, the total Hamiltonian for $\nMode$ harmonic oscillators interacting through the Coulomb interaction is given by
\begin{align}
    \label{eqApp:bare_Hamiltonian}
    H&=\sum_j\left(\frac{\tilde{p}_j^2}{2m} + \frac{m\tilde{\omega}_0^2 \tilde{x}_j^2}{2}\right) \notag\\
    &\hspace{5mm}+\sum_{j>k}\left(\frac{-m\tilde{\omega}_0\tilde{\kappa}_{j,k}}{2}(\tilde{x}_j^2 + \tilde{x}_k^2) + \frac{\hbar\tilde{\kappa}_{j,k}}{x_0^2}\tilde{x}_j\tilde{x}_k\right),
\end{align}
where $\tilde{\kappa}_{j,k}=e^2/(4\pi\epsilon_0 d_{j,k}^3m\tilde{\omega}_0)$ and $x_0=\sqrt{\hbar/(m\omega_0)}$.
As can be seen in Eq.~\eqref{eqApp:bare_Hamiltonian}, the first term of the Coulomb interaction shifts the oscillation frequency, while the second term mixes the different modes.
Because the frequency shift depends on $j$ as represented by $\sum_k\tilde{\kappa}_{j,k}$, $\PLJ$ should also depends on $j$, which induces additional complexity.

Therefore, we replace $\tilde{\omega}_0$ by the $j$-dependent frequency $\tilde{\omega}_{j}$ in a manner that the shifted oscillation frequency becomes $j$ independent.
First, the Hamiltonian can be written as
\begin{align}
    \label{eqApp:bare_Hamiltonian_j-dependent}
    H&=\sum_j\left(\frac{\tilde{p}_j^2}{2m} + \frac{m\tilde{\omega}_{j}^2 \tilde{x}_j^2}{2}
    +\frac{-m\sum_{k\neq j}\tilde{\omega}_{j,k}\tilde{\kappa}_{j,k}}{2}\tilde{x}_j^2 \right) \notag\\
    &\hspace{5mm}+\sum_{j>k}\frac{\hbar\tilde{\kappa}_{j,k}}{x_{0,j}x_{0,k}}\tilde{x}_j\tilde{x}_k,
\end{align}
$x_{0,j}=\sqrt{\hbar/(m\tilde{\omega}_{j})}$, and $\tilde{\kappa}_{j,k}$ is redefined as $\tilde{\kappa}_{j,k}=e^2/(4\pi\epsilon_0 d_{j,k}^3m\tilde{\omega}_{j,k})$ with $\tilde{\omega}_{j,k}=\sqrt{\tilde{\omega}_{j}\tilde{\omega}_{k}}$.
We eliminate the $j$ dependence of the oscillation frequency by requiring
\begin{align}
    \omega_0^2 &= \tilde{\omega}_{j}^2 - \sum_{k\neq j}\tilde{\omega}_{j,k}\tilde{\kappa}_{j,k} \notag\\
    &=\tilde{\omega}_{j}^2 - \sum_{k\neq j}\frac{e^2}{4\pi\epsilon_0d_{j,k}^3m}. \label{eqApp:omega0}
\end{align}
Therefore, for a given $\omega_0$, by tuning the harmonic potential of each mode so that the bare oscillation frequency $\tilde{\omega}_j$ satisfies Eq.~\eqref{eqApp:omega0}, Eq. \eqref{eqApp:bare_Hamiltonian_j-dependent} can be rewritten as
\begin{align}
    \label{eqApp:bare_Hamiltonian_j_dependence_in_eliminated}
    H&=\sum_j\left(\frac{\tilde{p}_j^2}{2m} + \frac{m\omega_{0}^2 \tilde{x}_j^2}{2} \right) +\sum_{j>k}\frac{\hbar\tilde{\kappa}_{j,k}}{x_{0,j}x_{0,k}}\tilde{x}_j\tilde{x}_k.
\end{align}

Then, by introducing the scaled coordinate and momentum defined, respectively, as $x=\tilde{x}/x_0$ and $p=\tilde{p}/p_0$ with  $x_0=\sqrt{\hbar/(m\omega_0)}$ and $p_0=\sqrt{\hbar m\omega_0}$, the Hamiltonian is given by
\begin{align}
    \label{eqApp:Hamiltonian_scaled_coordinate}
    H=\sum_j\frac{\hbar\omega_0}{2}(p_j^2 + x_j^2) + \sum_{j>k}\hbar\kappa_{j,k}x_jx_k.
\end{align}
By using the relations $x_j=(a^{\dagger}_j + a_j)/\sqrt{2}$ and $p_j=\ii(a^{\dagger}_j - a_j)/\sqrt{2}$, the Hamiltonian in the Fock basis is obtained as
\begin{align}
    \label{eqApp:Hamiltonian_Fock}
    H=&\sum_j\hbar\omega_0\left(a_j^{\dagger}a_j + \frac{1}{2}\right) \notag\\
    &+\sum_{j>k}\frac{\hbar\kappa_{j,k}}{2}\left(a_j^{\dagger}+a_j\right)\left(a_k^{\dagger}+a_k\right).
\end{align}
When the rotating-wave approximation is applied, the terms which do not preserve the number of phonons vanish and the resultant Hamiltonian becomes
\begin{align}
    \label{eqApp:Hamiltonian_Fock_RWA}
    H&=H_0+\sum_{j>k} H_{j,k} \notag\\
    &=\sum_j\hbar\omega_0\left(a_j^{\dagger}a_j + \frac{1}{2}\right)
    +\sum_{j>k}\frac{\hbar\kappa_{j,k}}{2}\left(a_j^{\dagger}a_k+a_ja_k^{\dagger}\right).
\end{align}
\section{Phase shift and \bs}\label{app:relations}
The idea of the DD is based on the following relations;
\begin{align}
    e^{\ii\phi a^{\dagger}a} a e^{-\ii\phi a^{\dagger}a} = e^{-\ii\phi}a, \\
    e^{\ii\phi a^{\dagger}a} a^{\dagger} e^{-\ii\phi a^{\dagger}a} = e^{\ii\phi}a^{\dagger}.
\end{align}
When the two-mode \bs is defined as $\BTM_{1,0}(\theta)=\exp(-\ii\theta(a_1^{\dagger}a_0 + a_1a_0^{\dagger}))$ and the phase shift is defined as $P_1(\phi)=\exp(-\ii\phi a_1^{\dagger}a_1)$, we derive
\begin{align}
    \label{eqApp:phase_shift_and_beamsplitter}
    P_1^{\dagger}(\phi) \BTM_{1,0}(\theta) P_1(\phi)
    =\exp\left(-\ii\theta \left(e^{\ii\phi}a_1^{\dagger}a_0 + e^{-\ii\phi}a_1a_0^{\dagger}\right)\right),
\end{align}
and the similar equations for $P_0(\phi)$ and $P_0^{\dagger}(\phi)$.
Therefore, by denoting the $\pi$-phase shift as $\Ppi_j$ and using $\Ppi_j=\Ppi_j^{\dagger}$, we derive
\begin{align}
    \label{eqApp:basic_relation_DD}
    \Ppi_j \BTM_{1,0}(\theta) \Ppi_j = \BTM_{1,0}(-\theta), \, j = 0~\mathrm{or}~1.
\end{align}
Because the phonon-hopping Hamiltonian satisfies
\begin{align}
    \Ppi_{\xi} H_{j,k} \Ppi_{\xi} = - H_{j,k}, \, \xi=j~\mathrm{or}~k,
\end{align}
Eq.~\eqref{eqApp:basic_relation_DD} can be extended to the $\nMode$-mode entangling gate $B_{\nMode}(\btheta)=\exp(-\ii\tau\sum_{j>k}H_{j,k}/\hbar)=\exp(-\ii\sum_{j>k}\theta_{j,k}(a_j^{\dagger}a_k + a_j a_k^{\dagger}))$ as
\begin{align}
    \label{eqApp:multi-mode_DD}
    &\Ppi_j B_{\nMode}(\btheta) \Ppi_j \notag\\
    &\hspace{10mm}= \exp\left(-\ii\sum_{j'>k}(-1)^{\delta_{j,j'}}\theta_{j',k}\left(a_{j'}^{\dagger}a_k + a_{j'} a_k^{\dagger}\right)\right),
\end{align}
where $\delta_{j,j'}$ is the Kronecker delta.
Consequently, because $\Ppi_j$ flips the sign of the phonon hopping terms containing $a_j$ or $a_j^{\dagger}$, 
Eqs.~\eqref{eq:3mode_relation_P1_and_B+-} and \eqref{eq:3mode_relation_P2_and_B+-} are derived.

\section{Time-dependent harmonic oscillator}\label{app:Lau&James}
A time-dependent harmonic oscillator (TDHO) is defined by the following Hamiltonian
\begin{align}
    \label{eqApp:Hamiltonian_TDHO}
     H=\frac{\tilde{p}^2}{2m} + \frac{1}{2}m\omega^2(t) \tilde{x}^2,
\end{align}
which can be decomposed into $h_0=\tilde{p}^2/(2m) + m\omega_0^2 \tilde{x}^2 / 2$ and the time-dependent modulation 
\begin{align}
    \HLJ(t) &= H-h_0
    = \frac{1}{2} m\Omega^2(t) \tilde{x}^2 \notag\\
    &= \frac{\hbar\Omega^2(t)}{2\omega_0} x^2 \notag\\
    &= \frac{\hbar\Omega^2(t)}{4\omega_0} (a^{\dagger} + a)^2,
\end{align}
where $\Omega^2(t)=\omega^2(t) - \omega_0^2$.
The solution of $\ii\partial_t\ket{\psi}=H\ket{\psi}$ is known to be given by a closed form \cite{LewisRiesenfeld1969, Lohe2008}.
When the initial state at $t=0$ is the $n$-th eigenstate $\varphi_n$ of $h_0$, the solution at $t$ is given in the scaled coordinate by
\begin{widetext}
    \begin{equation}
        \psi = \sqrt{\frac{1}{2^n n!\sqrt{\pi}b x_0}}
        \mathcal{H}_n\left(\frac{x}{b}\right)
        \exp\left(-\frac{1}{2}\left(\frac{x}{b}\right)^2\right)
        \exp\left(\frac{\ii\dot{b}}{2b\omega_0}x^2\right)
        \exp\left(-\ii\left(n+\frac{1}{2}\right)\omega_0\int_0^{\tphi}\frac{dt'}{b^2}\right),
    \end{equation}
\end{widetext}
where $\mathcal{H}_n$ is the Hermite polynomial and $b$ is related to $\omega(t)$ by
\begin{align}
    \label{eqApp:b_omegat_relation}
    \ddot{b}+\omega^2(t)b - \frac{\omega_0^2}{b^3} = 0.
\end{align}
When the boundary conditions, $b(t\leq0)=b(t\geq\tphi)=1$ and $\omega(0)=\omega(\tphi)=\omega_0$, are satisfied, 
$\psi(t\geq\tphi)$ coincides with $\varphi_n$ 
except for the $n$-dependent phase factor. 
In the interaction picture with respect to $h_0$, $\psi$ acquires the phase $\exp(-\ii(n+1/2)\phi)$ at $t\geq\tphi$, where $\phi$ is defined by Eq.~\eqref{eq:phase_shift_LJ}.

\begin{figure}[htbp]
\begin{center}
\includegraphics[width=8.5cm]{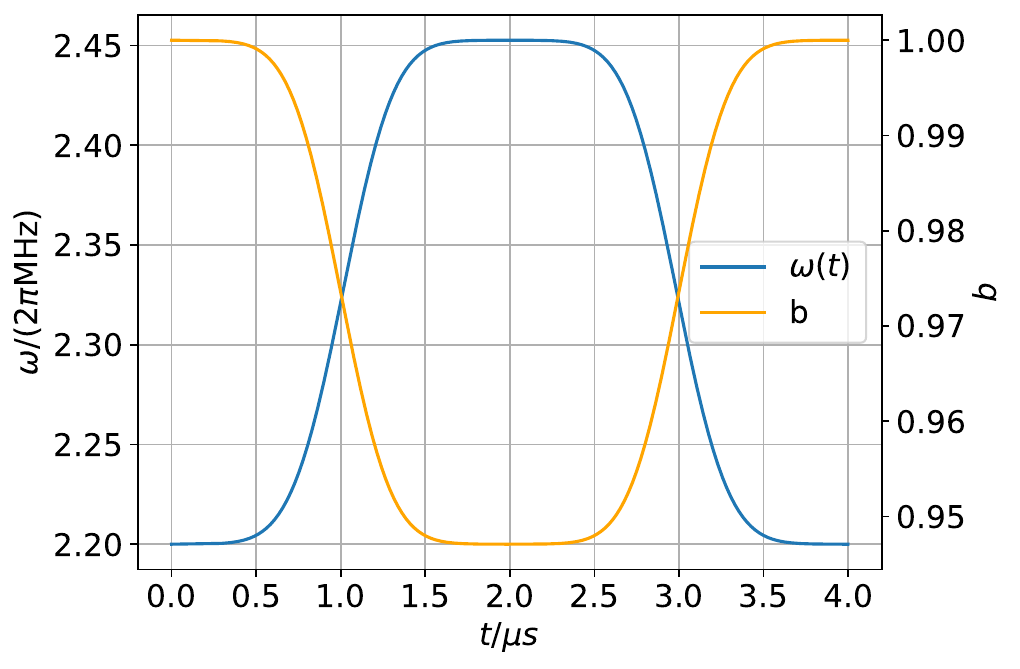}
\caption{The time-dependent frequency $\omega(t)$ and the $b$ function for the long pulse ($\tphi=4.0\,\mus$), with $\Tud=4.4T_0=2.0\,\mus$ and $k=0.0529$.}
\label{figApp:omega_t_and_b}
\end{center}
\end{figure}

The $b$ function defined by Eq.~\eqref{eq:b_errorfuncion} can be specified by the duration $\tphi$, the ramp-up and ramp-down time $\Tud$, the strength $k$, and the width of the error function $\sigma$, which we fix to $\sigma=6$.
In Sec. \ref{sec:Results}, we considered the short ($\tphi=2.2T_0=1.0\,\mus$) and long ($\tphi=8.8T_0=4.0\,\mus$) pulses, where $T_0=2\pi/\omega_0$ = 0.455$\,\mus$.
In order to obtain the $\pi$-phase shift $\PLJ$, we impose $\phi=\pi$ and determine the sets of the parameters to be $\{\Tud=2.0T_0,\,k=0.1636\}$ for the short pulse and $\{\Tud=4.4T_0,\,k=0.0529\}$ for the long pulse.
The $b$ function and $\omega(t)$ for the long pulse are shown in Fig.~\ref{figApp:omega_t_and_b}.

\begin{figure}[htbp]
\begin{center}
\includegraphics[width=8.5cm]{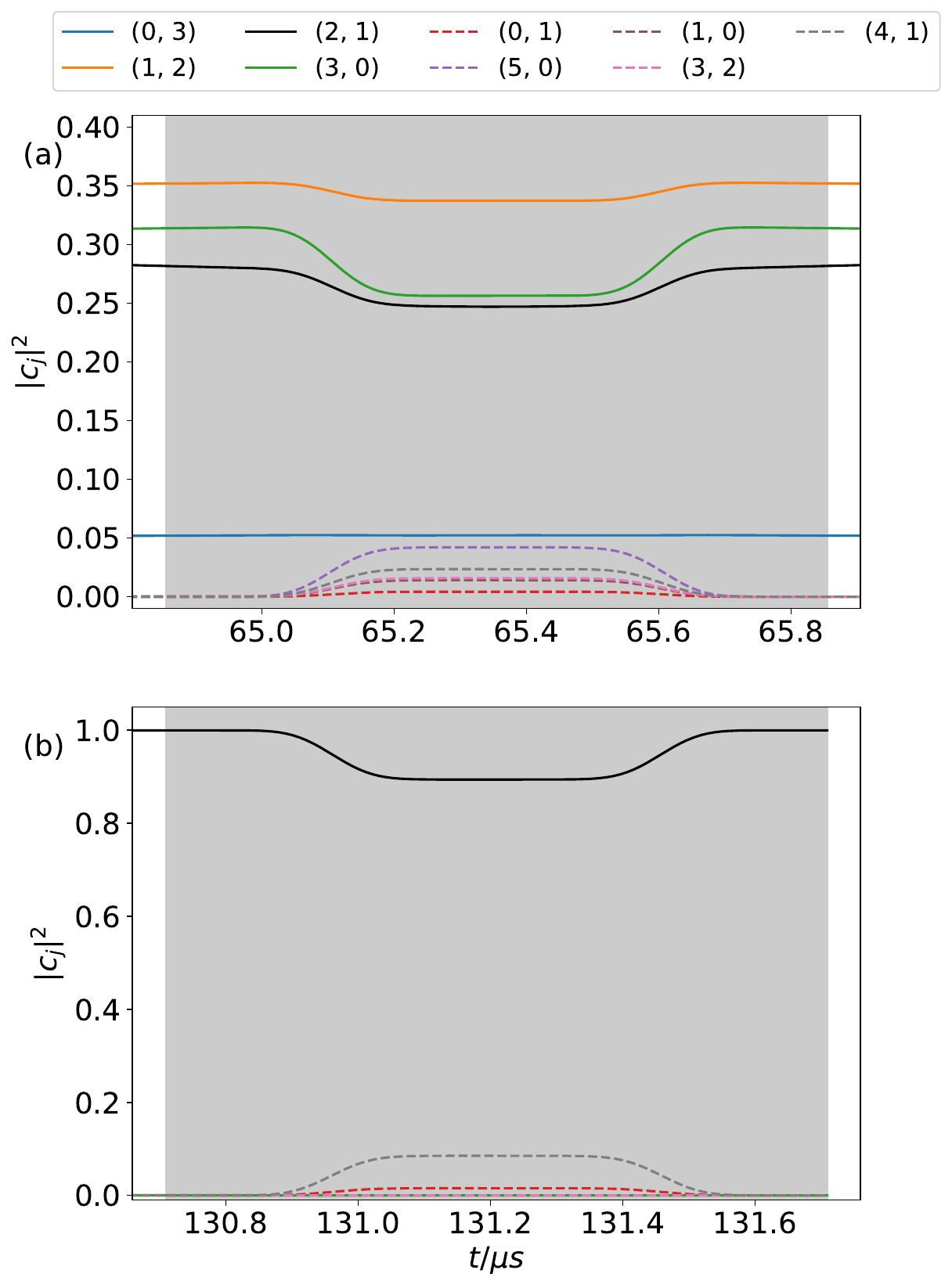}
\caption{The populations during the (a) first and (b) second applications of $\PLJ$ with $\tphi=1\,\mus$ and $\ket{\psi_0}=\ket{2,\,1}$.
In addition to the states (solid line) shown in Fig.~\ref{fig:2mode_C3PO_d27.6T4andT1}(b), 
the states (dashed line ) which do not preserve the phonon number are also shown if their population exceeds 1\% during the simulation.}
\label{figApp:zoom}
\end{center}
\end{figure}

During the application of $\PLJ$, the phonon number is not necessarily preserved.
For example, during the application of $\PLJ$ with $\tphi=1.0\,\mus$ in Fig.~\ref{fig:2mode_C3PO_d27.6T4andT1}(b), the populations of the phonon-number-preserving states decrease, while the states which do not preserve the phonon number are populated as shown in Fig.~\ref{figApp:zoom}.

\section{Electric pulse to realize the time-dependent harmonic oscillator}\label{app:electric_pulse}
When we apply an oscillating voltage $V(t)=U_0 + V_0 \cos(\Omega_r t)$ between diagonal electrodes of a linear trap, the quadrupole potential is created near the trap axis $\tilde{z}$ \cite{Wineland1998}, where $U_0$ and $V_0$ are the amplitudes of the DC and RF voltages, respectively, and $\Omega_r$ is the RF frequency.
The motion of the ion along the radial directions, $\tilde{x}$ and $\tilde{y}$, can be separated into the secular and micro motions when $|a_{\alpha}|,\,|q_{\alpha}|\ll 1\,(\alpha=\tilde{x},\,\tilde{y})$ are satisfied for $a_{\tilde{x}}=-a_{\tilde{y}}=4eU_0/(m\Omega_r^2r_0^2)$ and $q_{\tilde{x}} = -q_{\tilde{y}} = 2eV_0/(m\Omega_r^2r_0^2)$, where $r_0$ is the distance from the trap axis to the electrodes.
The effective potential for the secular motion is given by
\begin{align}
    \Phi_{\mathrm{eff}} = \frac{m}{2e}\left(
    \omega_{0\tilde{x}}^2\tilde{x}^2 + \omega_{0\tilde{y}}^2\tilde{y}^2 
    \right),
\end{align}
where $\omega_{0\alpha}=\sqrt{a_{\alpha}+q_{\alpha}^2/2}\,\Omega_r/2$.

In addition, we need to apply a DC voltage to the end electrodes to provide confinement along the $\tilde{z}$ direction.
Therefore, the three-dimensional harmonic potential near the trap axis is given as
\begin{align}
    \label{eqApp:eff+DC_potential}
    \Phi_{\mathrm{eff}} + \Phi_{\tilde{z}}
    = \frac{m}{2e}\left(
    \omega_{\tilde{x}}^2\tilde{x}^2 + \omega_{\tilde{y}}^2\tilde{y}^2 + \omega_{\tilde{z}}^2\tilde{z}^2 
    \right),
\end{align}
where $\omega_{\alpha}=\sqrt{\omega_{0\alpha}^2 - \omega_{\tilde{z}}^2/2}$ for $\alpha=\tilde{x},\tilde{y}$.

In order to realize the time-dependent frequency $\omega(t)$, we modulate $U_0$ or $V_0$.
When we modulate the DC amplitude while fixing the RF amplitude,  $\omega_{\alpha}$ at $t$ is given as
\begin{align}
    \omega_{\alpha}(t)^2 = \frac{eU_0(t)}{mr_0^2} + \frac{q_{\alpha}^2\Omega_r^2}{8} - \frac{\omega_{\tilde{z}}^2}{2}.
\end{align}
By denoting $\omega_{\alpha}(t)$ as $\omega(t)$, we obtain the time-dependent DC amplitude for a given $\omega(t)$ as
\begin{align}
    \label{eqApp:U0_and_omega}
    U_0(t) = \frac{mr_0^2}{e} \left(\omega^2(t) + \frac{\omega_{\tilde{z}}^2}{2} -\frac{q_{\alpha}^2\Omega_r^2}{8} \right).
\end{align}
Similarly, when we fix the DC amplitude, the time-dependent RF amplitude is given by
\begin{align}
    \label{eqApp:V0_and_omega}
    V_0(t) = \frac{\sqrt{2}m\Omega_r r_0^2}{e} \sqrt{\omega^2(t) + \frac{\omega_{\tilde{z}}^2}{2} - \frac{a_{\alpha}\Omega_r^2}{4}}.
\end{align}


%

\end{document}